\documentclass[authoryear]{elsarticle}
\usepackage{amsthm}
\usepackage{amsmath}
\usepackage{amssymb}
\usepackage{epsfig}
\usepackage{epstopdf}
\usepackage{graphicx}
\usepackage{amsmath}
\usepackage{amssymb}
\usepackage{makeidx}
\usepackage{caption}
\usepackage{subcaption}
\usepackage{float}
\usepackage{hyperref}
\usepackage{color,soul}

\newcommand{\Fb}{\mathbf{F}}
\newcommand{\Lb}{\mathbf{L}}

\newcommand{\Feb}{\mathbf{F}^{e}}
\newcommand{\Fpb}{\mathbf{F}^{p}}
\newcommand{\Deb}{\mathbf{D}^{e}}
\newcommand{\Ceb}{\mathbf{C}^{e}}
\newcommand{\Dpb}{\mathbf{D}^{p}}
\newcommand{\Web}{\mathbf{W}^{e}}
\newcommand{\Wpb}{\mathbf{W}^{p}}
\newcommand{\Leb}{\mathbf{L}^{e}}
\newcommand{\Lpb}{\mathbf{L}^{p}}
\newcommand{\Eeb}{\mathbf{E}^{e}}
\newcommand{\nup}{\nu^p}
\newcommand{\Npb}{\mathbf{N}^{p}}
\newcommand{\Seb}{\mathbf{S}^{e}}
\newcommand{\Tb}{\mathbf{T}}
\newcommand{\Teb}{\mathbf{T}^{e}}
\newcommand{\Qb}{\mathbf{Q}}
\newcommand{\xb}{\mathbf{x}}
\newcommand{\yb}{\mathbf{y}}
\newcommand{\vb}{\mathbf{v}}
\newcommand{\chib}{\mathbf{\chi}}
\newcommand{\invFeb}{{\mathbf{F}^e}^{-1}}
\newcommand{\invFpb}{{\mathbf{F}^p}^{-1}}
\newcommand{\tildenup}{\tilde{\nu}^p}
\newcommand{\tildeLeb}{\tilde{\mathbf{L}}^{e}}

\begin{document}

\title{Two-temperature continuum model for viscoplasticity in metals based on fluctuation relation}

\author[IISc]{Shubhankar~Roy~Chowdhury}
\ead{shuvorc@civil.iisc.ernet.in}

\author[IISc]{Debasish~Roy\corref{cor1}}
\ead{royd@civil.iisc.ernet.in}

\author[TAMU]{J~N~Reddy}
\ead{jnreddy@tamu.edu}

\cortext[cor1]{Corresponding author}
\address[IISC]{Computational Mechanics Lab., Department of Civil Engineering, Indian Institute of Science, Bangalore 560012, India}
\address[TAMU]{Advanced Computational Mechanics Lab., Department of Mechanical Engineering, Texas A\&M University, College Station, Texas 77843-3123}

\begin{abstract}
A continuum plasticity model for metals is presented from considerations of non-equilibrium thermodynamics. Of specific interest is the application of a fluctuation relation that subsumes the second law of thermodynamics en route to deriving the evolution equations for the internal state variables. The modeling itself is accomplished in a two-temperature framework that appears naturally by considering the thermodynamic system to be composed of two weakly interacting subsystems, viz. a kinetic vibrational subsystem corresponding to the atomic lattice vibrations and a configurational subsystem of the slower degrees of freedom describing the motion of defects in a plastically deforming metal. When externally driven, the two subsystems, identified with their own temperatures, fall out of equilibrium. An apparently physical nature of the present model derives upon considering the dislocation density, which characterizes the configurational subsystem, as a state variable. The continuum model accommodates finite deformation and describes plastic deformation in a yield-free setup via the so-called microscopic force balance along with the more conventional macroscopic force balance. Though the theory here is essentially limited to face-centered cubic metals modelled with a single dislocation density as the internal variable, an extension to body-centered cubic metals and modelling with two internal variables with dislocation densities of two non-overlapping types are briefly touched upon.
\end{abstract}


\maketitle

\section{Introduction}
The physical processes underlying plastic or viscoplastic deformation in metals are extremely complex and inherently irreversible. Plastic deformation originates from the motion of microscopic defects, primarily the crystallographic slips caused by dislocations. Even though several other mechanisms including twining, grain boundary sliding, void growth etc. do influence plastic deformation, our present focus will solely be on dislocation motion and evolution as the predominant micromechanism driving metal plasticity. In the microscopic scale, the motion of dislocations and their interactions with each other as well as the lattice itself are highly nonlinear and temporally intermittent dynamical flows that render an inhomogeneous texture to spatial deformation fields with several characteristic dislocation pattern formations. Still, at the macroscale, the field appears smooth and homogeneous. Motion of dislocations through the lattice requires overcoming an energy barrier with the combined aid of the applied force and thermal fluctuations. The crystal lattice configuration, viz. face-centered cubic (FCC), body-centered cubic (BCC), hexagonal close-packed (HCP), plays an important role in deciding how the thermal activation affects the mechanical response and thus plastic deformations in metals differ depending on the crystal structure.

When driven by an external protocol, a crystalline material with dislocations falls out of equilibrium because of the self energy of dislocations. \citet{langer2010} argued that a macroscopic system undergoing plastic deformation might be interpreted as being composed of slow configurational degrees of freedom describing infrequent and intermittent atomic rearrangements responsible for the plastic flow and kinetic-vibrational degrees of freedom pertaining to the thermally induced vibrational motion of atoms. Consequently the system may be split into a configuration subsystem and a kinetic-vibration subsystem. The energy of the configurational subsystem, i.e. the self energy of dislocations, along with its own entropy defines a new temperature -- the configurational or effective temperature, which must be distinguished from the thermal or kinetic vibrational temperature. As plastic deformation progresses at the expense of external work, the effective temperature evolves differently from the thermal temperature and establishes a heat current from the configurational subsystem to the other. Consequently a large fraction of the external work that generates dislocations and keeps them in motion, dissipates as heat. This makes evident the highly irreversible and dissipative character of plastic deformation and also suggests the configurational rearrangements to be far from equilibrium.

A desirable, if challenging, goal in continuum mechanics is to develop predictive viscoplasticity models for the complex phenomena of dislocation evolution and motion. The models should be applicable to a wide range of temperature and strain rates as the deformation response of metals depend, to a large extent, on these two parameters. More importantly, several engineering applications of metals at high strain rates, e.g. high speed machining, impact on armor systems, metal forming etc., demand an understanding of the underlying micromechanisms of plastic deformation and their exploitation within the predictive model in order to optimize or enhance the design and operations of such systems. Huge research efforts, spanning the last few decades and aimed at developing constitutive models -- mostly phenomenological -- of plasticity or viscoplasticity, have perhaps not led to a proportionate increment in the physical understanding that might have culminated in a universally acceptable theory. Many such models are posited on the hypothesis of local equilibrium thermodynamics, wherein internal state variables are utilized to characterize the irreversible processes of viscoplasticity. In parallel, quite a few reported efforts at developing a physically based theory for viscoplastic deformation also do exist. However, rather than exploiting the thermodynamic principles, many such models often postulate the constitutive relations in terms of the evolution equations of dislocation densities. These equations explicitly contain terms describing several microscopic phenomena, e.g., dislocation multiplication, annihilation, grain size effect, effect of cell boundaries, twining, interactions with the lattice and so on. Some of the physically based models also bring in refinements by making a distinction among different types of dislocations - viz. mobile and immobile, statistically stored and geometrically necessary dislocations. Separate evolution equations for different dislocation types are then put forth whilst accounting for their mutual interactions in some way.

It is well known that the notions of temperature, heat, energy etc. have their moorings in thermodynamics. However thermodynamics, in its original form, merely provides for a basis encapsulating the principles of equilibrium that a macroscopic system is often believed to follow. Macroscopic work-heat relation is one such example.
Founded on Gibbs' ensemble formalism, equilibrium statistical mechanics made it possible to bridge widely separated scales, i.e. micro to macro. For example, it identified the thermodynamic temperature with the vibrational kinetic energy of the constituent particles, associated entropy with the probability distribution of microstates that conform to the specified macroscopic conditions and so on. Unfortunately, away from equilibrium, not only do most of these identifications fail to hold, definitions of quantities such as entropy and temperature also become ambiguous. To be sure, there exists a natural extension of thermodynamics to close-to-equilibrium systems. This extension is based on the local equilibrium hypothesis: the local and instantaneous relations between the thermodynamic quantities in a system out of equilibrium are the same as those for a uniform system in equilibrium. Representing state variables as spatial continuum fields hold only when the length scale over which the continuum fields vary appreciably is much larger than the microscopic length scale. Similarly, the time dependence of the field can be justified when changes occur sufficiently slowly vis-a-vis the macroscopic scale so that at each instant the thermodynamic system underlying each continuum particle gets sufficient time to equilibrate. Because of the usual disparity between macroscopic and microscopic scales, many systems do conform to the notion of local equilibrium. There are however situations galore where this hypothesis is invalid and the process of viscoplastic deformation of metals might very well be one of them. Therefore, a physical understanding of metal plasticity must be based on a confluence of ideas rooted in the geometrical aspects of plastic deformation and a thermodynamic theory applicable in far-from-equilibrium conditions.

Unlike the classical irreversible thermodynamics woven around the local equilibrium hypothesis, there is no natural extension of the equilibrium theory to generic non-equilibrium conditions. There are many schools of contemporary non-equilibrium theories: rational thermodynamics \citep{truesdell2012rational}, extended irreversible thermodynamics \citep{jou1996extended}, rational extended thermodynamics \citep{muller2013rational} and GENERIC \citep{ottinger2005beyond} to cite a few. These theories differ from one other in various aspects, e.g. one may consider temperature and entropy as primitive concepts when the other derives them; dissipation inequality in global or local form may be postulated or based on experimental facts; relation between the heat flux density and entropy flux density may be universal in one and material-dependent in another. Even as most of these theories employ the second law of thermodynamics in form of a dissipation inequality, they again differ in the specific way the inequality is exploited. Many of them however lack a foundation in statistical physics, its extension to the non-equilibrium case itself being not firmly established. There are many non-equilibrium statistical operator methods available for the statistical description of non-equilibrium process. One formalism deserves a special mention. It is the predictive statistical mechanics based on information theory and Bayesian inference \citep{jaynes1957information,luzzi2013predictive}, which combines mechanical and statistical aspects with the macroscopic observables and extends Gibbs' equilibrium ensemble formalism to systems arbitrarily away from equilibrium.

In the theories above cited, with or without a statistical basis, the second law of thermodynamics plays a crucial role. In continuum theory of viscoplasticity, the second law acts as a restriction on the material constitution and helps postulating evolution laws of the state variables. Beyond the pale of the inequality of the second law, a class of equalities, all of fairly recent origin and applicable even far away from equilibrium, exists. Integral fluctuation relations \citep{seifert2005entropy,speck2005integral}, detailed fluctuation relations \citep{evans1993probability,gallavotti1995dynamical,kurchan1998fluctuation,maes1999fluctuation}, work and free-energy equality \citep{jarzynski1997nonequilibrium,crooks1999entropy} are a few examples. Indeed, they generalize the statement of the second law by relating the notions of macroscopic entropy, work, free-energy, entropy production etc. to the microscopic path space. Clearly, applying an equality of this type in place of the second law would impose stricter restrictions on the constitutive modeling within a continuum viscoplasticity model. Addressing this point is one of the novel features of this article.

For our formulation on viscoplasticity in metals, we presently consider dislocation motion and evolution as the sole factor responsible for plastic deformation. Following \citet{langer2010} and  \citet{langer2015statistical}, dislocations are assumed to reside in the configurational subsystem where changes due to plastic deformation occur much more slowly compared to the kinetic vibrational motion in atoms. Supplementing the usual forms of force balance, energy balance etc., a variant of the integral fluctuation relation is used to derive the constitutive restrictions and the evolution equation for dislocation density. Here again, a continuum formulation for viscoplasticity can be attained in several ways, viz. consistency viscoplasticity, Perzyna type models \citep[see][for both the formats]{lubliner2008plasticity,simo2006computational}, yield free theories \citep{gurtin2010mechanics, bodner1975constitutive} etc. We construct the present theory in a yield-free setup fashioned after \citet{gurtin2010mechanics}.

The rest of the paper is organised as follows. In  section \ref{sec_Fluctuation}, a brief account on the fluctuation theorem, specifically the integral form of it, is given. The two-temperature framework for viscoplasticity and application of the fluctuation theorem in developing a theory of viscoplasticity is described in section \ref{sec_Fluctuation_plast}. In these two sections, considerations are restricted only to the case of homogeneous fields. However, based on the developments of section \ref{sec_Fluctuation_plast}, a full fledged formulation of continuum viscoplasticity applicable to spatially inhomogeneous fields is undertaken in section \ref{sec_Cont_form}, wherein the finite deformation kinematics, balance of macroscopic and microscopic forces, and constitutive relations are described. The constitutive relations, derived in section \ref{sec_Cont_form}, are applicable for FCC metals and written using dislocation density as a single internal variable. Brief accounts on how to extend the theory to BCC metals and include different types of dislocation densities in the constitutive description are presented in section \ref{sec_extension}. This is followed by a few elementary numerical experiments in section \ref{sec_numerical}. Finally, the work is concluded in section \ref{sec_conclusion}.

\section{Fluctuation relation}\label{sec_Fluctuation}
Systems undergoing diverse thermodynamic processes, e.g. heat transfer, chemical reaction, mass diffusion and plastic deformation which is of central importance here, obey the first and second laws of thermodynamics. While the first law describes a balance among the internal energy increase, external work done and heat input, the second law relates to the arrow of time, providing directionality to the processes with the prescription that entropy production in a system is never negative. The last statement must however be subjected to a nuanced interpretation in line with the principles of statistical mechanics, which assert that second law holds only in a statistically averaged sense and hence could be violated over a subsets, perhaps of low probability measures, in the associated $\sigma$-algebra. The statement of the second law also led to Loschmidt’s reversibility paradox: even if the underlying microscopic mechanics is time-reversal invariant, the entropy function increases in forward time making the macroscopic process irreversible. In resolving this paradox, fluctuation theorems(FTs) have been derived over the last two decades. The FT refers to a symmetry relation of the following type.
\begin{equation}\label{eq_fluc1}
\frac{P_F\left(\Sigma\right)}{P_R\left(-\Sigma\right)} = \exp\left(\Sigma\right)
\end{equation}
Here $P_F\left(\Sigma\right)$ is the probability distribution of the entropy production, $\Sigma$, over a specified time interval, and $P_R\left(-\Sigma\right)$ is the probability distribution of the entropy production when the system is driven in a time-reversed manner.
The definition of entropy production depends on the dynamics used to model the evolution of the system. The FT has been established under reasonable definitions of entropy generation for a variety of physical situations modelled using deterministic or stochastic equations of motion. To elaborate, Eq. \eqref{eq_fluc1} says that observing positive entropy generation becomes exponentially more likely than the negative entropy generation as the system size grows. However, for sufficiently small systems or over very small time intervals, entropy production could be negative and the FT quantifies this result. For details on the derivation of the FT, see \citet{evans1993probability,gallavotti1995dynamical,kurchan1998fluctuation,maes1999fluctuation,crooks1999excursions} and the references therein.

The FT encapsulates an exact expression applicable to any system for which the entropy production function is well defined. Considering microscopic entropy generation as a measure of the thermodynamic reversibility of a path-space trajectory, an expression for entropy generation, in the framework of stochastic thermodynamics, can be given by the following.
\begin{equation}\label{eq_fluc2}
 \Sigma := \ln\frac{P\left(X_F\right)}{\tilde{P}\left(X_R\right)} = \ln\frac{T\big(x\left(\tau\right)|x_0\big)p_0(x_0)}{\tilde{T}\big(\tilde{x}\left(\tau\right)|\tilde{x}_0\big)p_1(\tilde{x}_0)}
\end{equation}
The first equality is with the logarithm of the ratio of probability of observing an arbitrary system trajectory $X_F$ to that of observing the time-reversed trajectory $X_R$ in the same ensemble of trajectories. The reverse trajectory may be generated from the forward as $\tilde{x}(\tau):= x(t-\tau)$. Given the initial point $x_0$ and the terminal point $x_t$ of a trajectory, the following holds.
$$ x_0 := x(0)=\tilde{x}(t)=:\tilde{x}_t  \quad \text{and} \quad  x_t := x(t)=\tilde{x}(0)=: \tilde{x}_0 $$
Assuming the processes as Markovian and given the initial and terminal points, $T\big(x\left(\tau\right)|x_0\big)$ and $\tilde{T}\big(\tilde{x}\left(\tau\right)|\tilde{x}_0\big)$ respectively refer to the transition probabilities of forward and backward paths. $p_0(x_0)$ and $p_1(\tilde{x}_0) = p_1(x_t)$ denote normalized distributions for initial and terminal values. With the definition of entropy production given in Eq. \eqref{eq_fluc2}, one can readily derive the following \citep[cf.][]{seifert2005entropy}.
\begin{equation}\label{eq_fluc3}
\begin{split}
 \langle\exp\left(-\Sigma\right)\rangle = \int P\left(X_F\right)\exp\left(-\Sigma\right) dX_F &= \int P\left(X_F\right)\frac{\tilde{P}\left(X_R\right)}{P\left(X_F\right)} dX_F \\ &= \int {\tilde{P}\left(X_R\right)}dX_R = 1
\end{split}
\end{equation}
Eq. \eqref{eq_fluc3} basically represents another form of the fluctuation relation, viz. the integral fluctuation theorem (IFT). $\langle\cdot\rangle$ denotes ensemble averaging in path space.

From IFT one may obtain the following relations.
\begin{equation}\label{eq_fluc4}
 \exp(0) =1 = \langle\exp\left(-\Sigma\right)\rangle \leq \exp\left(-\langle\Sigma\rangle\right)
\end{equation}
Here the last relation follows upon application of Jensen's inequality. Eq. \eqref{eq_fluc4} then trivially yields $\langle\Sigma\rangle \geq 0$, which is basically the statement of second law of thermodynamics: generation of the macroscopic entropy, which is obtainable as the path space average of microscopic entropy generation, is always non-negative. It is, therefore, evident that the IFT constitutes a generalization over the second law. Specifically, its representation in the form of an equality motivates us to investigate its implications in formulating a viscoplasticity theory for metals.

One may, in fact, go further and consider $\exp\left(-\Sigma\right)$ as an exponential martingale \citep{revuz2013continuous}. This interpretation should envisage entropy production as a uniquely expressible submartingale admitting a decomposition based on a strictly increasing deterministic component (typically a quadratic variation process) and a martingale with zero mean. While the strictly increasing component is typically passed off as the entropy production in the literature, we believe that the martingale component could have a crucial role in predicting bifurcation and instability in the material behaviour. This approach is however kept beyond the scope of this work.

\section{Fluctuation relation applied to viscoplasticity}\label{sec_Fluctuation_plast}
Under external driving, metals may deform plastically. Viscoplastic deformation of metals is a complex phenomenon that originates from highly nonlinear interactions of microscopic defects. Modelling viscoplasticity, therefore, demands material description through reliable constitutive models capable of describing the dynamic response of the system at large strain, high strain rate, under large temperature gradient etc. With plenty of empirical models for metal plasticity at hand, phenomenological theories appear to predominate this field of research. Johnson-Cook model \citep{johnson1983constitutive} is one such popular viscoplasticity model. While parameters in many such models might perhaps be tweaked to comply with the experimental observations over a broad strain rate regime, from small to very large, this cannot detract from the universal appeal of a physical theory that combines the geometry of micromechanisms responsible for viscoplastic deformation with the non-equilibrium features of the associated thermodynamics. Physically based theories for metal plasticity, for instance, write constitutive equations by relating the deformation micromechanics to the thermal activation of dislocations. Short range effects such as lattice friction and long range barriers-- interaction with grain boundaries, cell walls etc. are also considered. Models proposed by \citet{zerilli1987dislocation,aifantis1987physics,follansbee1988constitutive,klepaczko1996numerical,voyiadjis2005microstructural,gao2012constitutive} are a few examples in this category.

More recently, \citet{huang2009constitutive,langer2010,langer2015statistical,vinogradov2015irreversible} have developed plasticity theories based on considerations of irreversible thermodynamics and some concepts in statistical mechanics. Among these, \citet{langer2010} and \citet{langer2015statistical} take a very different view in describing plastic deformation of metals. We will adopt this interpretation in this work.

\subsection{The thermodynamic system}
Elasto-viscoplastic deformation of metals results from lattice distortion, and motion/rearrangement of defects through the lattice. The defect dynamics is coupled with the vibration of atoms in the lattice about their equilibrium positions. This vibration generates kinetic energy and manifests as thermodynamic temperature that can be measured by an ordinary thermometer. A thermodynamic system representing elasto-viscoplastic deformation process of metals should therefore include states capable of describing thermal, lattice distortion and defect motion aspects. However, the time scale in which thermal or kinetic vibration occurs is much smaller than that associated with the other processes, viz. atomic rearrangement, defect motion etc. Existence of such phenomena that occur in seemingly different time scales allows us to subdivide the thermodynamic system into two subsystems -- a \emph{kinetic-vibrational} subsystem and a \emph{configurational} subsystem. While the kinetic-vibrational subsystem includes the fast vibrating degrees of freedom, the configurational subsystem comprises of the slower degrees of freedom describing the evolving lattice defects. These two subsystems weakly interact with each other by heat transfer.

The viscoplastic response of metals with dissimilar crystal structures are characteristically different owing to the differences in the available slip systems. The type of crystal lattice also crucially affects the thermal activation mechanism and its consequent effect on the macromechanical response. Accordingly, in this section and the next, we focus on developing a theory applicable to metals with FCC crystal structure. Extension to metals with BCC structure will be briefly discussed in section \ref{sec_extension}.

Evolution of dislocations often associates to the dominant microscopic processes that underlie the macroscopic viscoplastic deformation in metals. Although dislocations of different kinds, e.g., mobile, immobile, statistically stored, geometrically necessary etc., exist in metals, it is possible, at least in some cases, to model plasticity introducing only one internal state variable that represents the density of dislocations without distinguishing among the types. We undertake this simple and convenient single internal variable modeling of dislocation mediated plasticity. In section \ref{sec_extension}, an extension based on two dislocation densities is commented on.

Following \citet{langer2010}, we consider as the thermodynamic system a slab of material of area $A$ and thickness $L$ in the plane of applied shear stress. We undertake a simpler case where no heat exchange occurs with the atmosphere and it is only the configurational heat, which gets generated during plastic deformation and flows from configurational subsystem to the kinetic-vibrational subsystem. From such a consideration, kinetic-vibrational subsystem acts as a heat bath to the configurational subsystem. Generalization is possible by considering yet another subsystem, representing the atmosphere, as the heat bath and allow for heat transfer from the kinetic-vibrational subsystem to it. See \citet{bouchbinder2009nonequilibrium} for a similar three-subsystem analysis for plasticity in amorphous materials. We, however, restrict the current formulation to the two-subsystem formalism. Also, to avoid the complexity of working with spatially varying fields, we consider the fields appertaining to plastic deformation as spatially homogeneous. Generalization to the inhomogeneous case and arbitrary geometry is considered in section \ref{sec_Cont_form}.

The total internal energy $U_t$ and the total entropy $S_t$ of the system are represented in the following way.
\begin{equation}\label{eq_FVP_5}
U_t = U_c\left(S_c,\rho\right) + U_k\left(S_k\right)
\end{equation}
\begin{equation}\label{eq_FVP_6}
S_t = S_c\left(U_c,\rho\right) + S_k\left(U_k\right)
\end{equation}
In Eq. \eqref{eq_FVP_5} and \eqref{eq_FVP_6}, component contributions from the two subsystems-- configurational and kinetic-vibrational, are written explicitly. $U_c$ and $U_k$ represent the configurational energy and kinetic-vibrational energy respectively. Similarly $S_c$ and $S_k$ denote the entropies of configurational and kinetic-vibrational subsystems. $\rho$ is the dislocation density expressed as dislocation length per unit volume. In describing $U_c$, we have excluded the elastic strain measure. This renders the system to behave as rigid plastic. Nevertheless, the conclusion arrived at regarding the dislocation density evolution remains unaffected. As noted, the kinetic-vibrational subsystem acts as a heat bath of temperature $\theta_k=k_BT=\partial{U_k}/\partial{S_k}$. $k_B$ is Boltzman's constant. For simplicity, we consider the bath temperature to remain fixed. Effects of both elastic distortion and evolving bath temperature will be considered in section \ref{sec_Cont_form}. The configurational subsystem assumes its own temperature $\theta_c$, called the effective temperature and given by the following.
\begin{equation}\label{eq_FVP_7}
\theta_c = \left(\frac{\partial{U_c}}{\partial{S_c}}\right)_\rho
\end{equation}
As the system is driven by external forces, $\theta_c$ starts evolving and falls out of equilibrium from the bath temperature $\theta_k$. This starts off a heat flux between the configurational subsystem and the bath and also plays an important role in the evolution of dislocations.

Assuming, the following additive decomposition of configurational energy and entropy, the contribution of dislocations can be separated out.
\begin{equation}\label{eq_FVP_8}
 U_c\left(S_c,\rho\right) = U_0\left(\rho\right) + U_1\left(S_c\right)
\end{equation}
\begin{equation}\label{eq_FVP_9}
S_c\left(U_c,\rho\right) = S_0\left(\rho\right) + S_1\left(U_c\right)
\end{equation}
While $U_0$ and $S_0$ designate configurational energy and entropy associated with the dislocations, $U_1$ and $S_1$ are the energy and the entropy of all the configurational degrees of freedom other than those pertaining to the dislocations. Dislocation energy $U_0$ is given by
\begin{equation}\label{eq_FVP_10}
U_0\left(\rho\right) = A \rho\, e_D,
\end{equation}
where $e_D$ is the energy per dislocation. Configurational entropy $S_0$ is determined by counting the number configurations for a given $U_c$ and $\rho$. This number may be arrived at by computing the number of ways lattice sites may be occupied by dislocations. For the slab of material considered, the number of dislocations may be found as $N_d = (\rho/L)V = \rho A$, where $V=AL$ is the total volume of the slab. This expression for $N_d$ trivially follows from the definitions of $\rho$ as length of dislocation per unit volume and $L$ as the characteristic length of dislocation. An expression for the number of lattice sites that can accommodate dislocations is given by $N_{ls} = V/(a^2L) = A/a^2$, where $a$ is a length of the order of atomic spacing and $a^2L$ may be visualized as a representative volume associated with each dislocation. The number of configurations, say $\Omega$, i.e, the number of ways $N_{ls}$ sites may be occupied by $N_d$ dislocations, is given by the following binomial coefficient.
\begin{equation}\label{eq_FVP_11}
\Omega = \begin{pmatrix}N_{ls} \\ N_d\end{pmatrix} = \frac{N_{ls}!}{N_d!\left(N_{ls}-N_d\right)!} = \frac{\left(N_{ls}\right)^{\underline{N_d}}}{N_d!}
\end{equation}
In Eq. \eqref{eq_FVP_11} $\left(N_{ls}\right)^{\underline{N_d}}$ denotes a falling factorial,
\begin{equation}\label{eq_FVP_12}
\left(N_{ls}\right)^{\underline{N_d}} = N_{ls}(N_{ls}-1)\cdots(N_{ls}-N_d+1)
\end{equation}
Since $1\ll N_d\ll N_{ls}$, $\left(N_{ls}\right)^{\underline{N_d}}$ can be approximated as $\left(N_{ls}\right)^{N_d}$.  With this, the expression for $S_0$ is given as follows.
\begin{equation}\label{eq_FVP_13}\begin{split}
S_0 = \ln\left(\Omega\right)=\ln\frac{\left(N_{ls}\right)^{N_d}}{N_d!} &= N_d\ln\left(N_{ls}\right)-\ln\left(N_d!\right) \\&\approx N_d\ln\left(N_{ls}\right)-N_d\ln\left(N_d\right)+N_d \\ &=- N_d\ln\frac{N_d}{N_{ls}} + N_d = -A\rho\ln\left(a^2\rho\right) + A\rho
\end{split}\end{equation}
Stirling's approximation, $\ln(N_d!) \approx N_d\ln(N_d) - N_d$ for large $N_d$, is made use of in writing the expression for $S_0$ in Eq. \eqref{eq_FVP_13}.

\subsection{First and second laws of thermodynamics}
With the definition of the thermodynamic system at hand, we can now apply the usual thermodynamic principles, viz. first and second laws. The first law describes the balance of the work input vis-a-vis the internal energy increase. In rate form, the first law is stated as
\begin{equation}\label{eq_FVP_14}
V\pi\nu_p = \dot{U_t} = \dot U_c + \dot U_k = \theta_c \dot S_c + \left(\frac{\partial U_c}{\partial\rho}\right)_{S_c}\dot\rho + \theta_k\dot S_k
\end{equation}
where $\pi$ is the so-called microforce responsible for viscoplastic deformation and $\nu_p$ the equivalent plastic strain rate. These quantities will be elaborated upon in section \ref{sec_Cont_form}.

The second law states that entropy production rate is never negative and, for the present system, this statement translates to
\begin{equation}\label{eq_FVP_15}
 \dot S_t = \dot S_c + \dot S_k \ge 0
\end{equation}
which may be written, after substituting for $\dot S_c$ from Eq. \eqref{eq_FVP_14}, in the following alternative form.
\begin{equation}\label{eq_FVP_16}
\frac{1}{\theta_c}\left[V\pi\nu_p - \left(\frac{\partial U_c}{\partial \rho}\right)_{S_c}\dot\rho\right] + \left(1-\frac{\theta_k}{\theta_c}\right)\dot S_k \ge 0
\end{equation}
Using Eq. \eqref{eq_FVP_8}, \eqref{eq_FVP_9}, \eqref{eq_FVP_10} and \eqref{eq_FVP_13},  one obtains
\begin{equation}\label{eq_FVP_17}\begin{split}
 \left(\frac{\partial U_c}{\partial \rho}\right)_{S_c} = \frac{\partial U_0}{\partial \rho} - \theta_c\frac{\partial S_0}{\partial \rho} = Ae_D + \theta_c A \ln\left(a^2\rho\right) = A\theta_c\left(\frac{e_D}{\theta_c}+\ln\left(a^2\rho\right)\right) \\=A\theta_c\ln\left(\frac{a^2\rho}{\exp\left(-\frac{e_D}{\theta_c}\right)}\right) \approx A\theta_c\left(\frac{\rho}{\rho_0}-1\right),
\end{split}\end{equation}
where $\rho_0 = \frac{1}{a^2}\exp\left(-\frac{e_D}{\theta_c}\right)$, and the approximation follows using the fact that $\ln(x)\approx x-1$ for $x\ll 1$. Eq. \eqref{eq_FVP_16}, can therefore be recast as
\begin{equation}\label{eq_FVP_18}
\frac{1}{\theta_c}\left[V\pi\nu_p + \left\{A\theta_c\left(1-\frac{\rho}{\rho_0}\right)\right\}\dot\rho\right] + \left(1-\frac{\theta_k}{\theta_c}\right)\dot S_k \ge 0
\end{equation}

It may be noted that, considering $S_c$ to be a function of $\theta_c$ and $\rho$, the first law as given in Eq. \eqref{eq_FVP_14} may also be written as the effective temperature evolution:
\begin{equation}\label{eq_FVP_19}
VC_{\text{eff}}\, d{\theta}_C =  d U_t - Ae_D d\rho - Qdt,
\end{equation}
where $Q=\theta_k\dot S_k$ is the heat transfer rate between configurational and kinetic-vibrational subsystems and $C_{\text{eff}}$ the specific heat of the configurational subsystem.

\subsection{Extension to path space and integral fluctuation theorem}
In going beyond the second law, we exploit the fluctuation theorem-- specifically the integral form of it as given in Eq. \eqref{eq_fluc3}. In order to investigate the restriction imposed by the IFT on the macroscopic constitutive relations or the evolution of state variables, we first need to extend the formulation to the path space. This can be done in a number of ways. One may start directly from the  microscale description of dislocation motion in terms of Langevin equations and develop the theory following the routes of stochastic thermodynamics \citep[see][]{seifert2005entropy}. This, however, is a very involved exercise. Therefore, it is reasonable to look for an alternative path-space description in terms of the associated probability distribution function. The MaxEnt formalism \citep{jaynes1957information} provides an elegant way to find out the state space probability distribution utilizing the given information about the macroscopic constrains, e.g. the macroscopic balance laws. See \citet{dewar2003information} for a derivation of fluctuation relation using the MaxEnt formalism. However, for present purposes, we implement an ad-hoc extension described below. Application of MaxEnt theory for two-temperature viscoplasticity will be undertaken elsewhere.

We consider the time varying thermodynamic states, i.e., the dislocation density, the effective  temperature, the total internal energy etc., as stochastic diffusion (Ito) processes, thus providing path-space representations of the corresponding states. We denote the stochastic states with superposed tilde, e.g., $\tilde \theta_c$, $\tilde \rho$, $\tilde{U}_t$, and so on. The stochastic states are assumed to admit the following decomposition.
\begin{equation}\label{eq_FVP_20}
\tilde \theta_c = \theta_c + \theta_{cf}, \quad \tilde \rho =  \rho +  \rho_f, \quad \tilde{U}_t = U_t + U_{tf}
\end{equation}
Here, $\theta_c$, $\rho$, $U_t $ denote the mean states and the quantities with subscript $f$, i.e., $\theta_{cf}$, $\rho_f$ and $U_{tf}$ represent random fluctuations about their mean values. The states above being strictly positive, one should, in principle, assume a decomposition of type: $\tilde X = X X_f$, where $X_f$ is a stochastic exponential. However, the additive decomposition given in Eq. \eqref{eq_FVP_20}, adopted here, is a reasonable approximation when the probability of fluctuations being large is  very small and it renders the subsequent derivation easier.

We now rewrite Eq. \eqref{eq_FVP_19} as a stochastic differential equation (SDE) as follows.
\begin{equation}\label{eq_FVP_21}
VC_{\text{eff}} \,d\tilde{\theta}_c =  d \tilde U_t - Ae_D d\tilde\rho - \delta\tilde Q,
\end{equation}
where $\delta\tilde Q$ refers to the incremental heat transfer between the subsystems and is given by
\begin{equation}\label{eq_FVP_22}
\delta\tilde Q = Q dt + dQ_f
\end{equation}
We may find an expression for the fluctuation of heat transfer in terms of fluctuations of other states by subtracting Eq. \eqref{eq_FVP_19} from Eq. \eqref{eq_FVP_22}.
\begin{equation}\label{eq_FVP_23}
dQ_f = dU_{tf} - VC_{\text{eff}}\,d\theta_{cf}-Ae_D d\rho_f
\end{equation}

Similarly, we write the path-space form of entropy production as follows.
\begin{equation}\label{eq_FVP_24}
d\left(\tilde{S}_c + \tilde{S}_k\right) = \frac{1}{\tilde{\theta}_c} d\tilde{U}_t - \frac{Ae_D}{\tilde{\theta}_c} d\tilde\rho - A\ln\left(a^2\tilde\rho\right)d\tilde\rho + \left(\frac{1}{\theta_k}-\frac{1}{\tilde{\theta}_c}\right)\delta\tilde Q
\end{equation}
One may contrast Eq. \eqref{eq_FVP_24}, describing the path-space entropy production, with the left hand side of \eqref{eq_FVP_16} or \eqref{eq_FVP_18}, the expression for mean entropy production rate.

The setup is now ready for applying the IFT:
\begin{equation}\label{eq_FVP_25}
\Big\langle \exp\left(-\int_{t_i}^{t_{i+1}}d\left(\tilde{S}_c + \tilde{S}_k\right)\right)\Big\rangle = 1
\end{equation}
To be specific, we consider the entropy generation over a small time interval $\left[t_i,t_{i+1}\right)$ and approximate it as
\begin{equation}\label{eq_FVP_26}\begin{split}
\int_{t_i}^{t_{i+1}}d\left(\tilde{S}_c + \tilde{S}_k\right) &\approx \frac{1}{\tilde{\theta}_{ci}} V\left(\pi\nup\right)_i \Delta t +\frac{1}{\tilde{\theta}_{ci}}\Delta U_{tf} - \frac{Ae_D}{\tilde{\theta}_{ci}} \dot{\rho}_i \Delta t -\frac{Ae_D}{\tilde{\theta}_{ci}}\Delta\rho_f\\ &- A\ln\left(a^2\tilde{\rho}_i\right)\dot{\rho}_i \Delta t - A\ln\left(a^2\tilde{\rho}_i\right)\Delta\rho_f \\& + \left(\frac{1}{\theta_k}-\frac{1}{\tilde{\theta}_{ci}}\right)Q_i\Delta t +\left(\frac{1}{\theta_k}-\frac{1}{\tilde{\theta}_{ci}}\right)\Delta Q_f,
\end{split}\end{equation}
where subscript $i$ indicates quantities evaluated at time $t_i$, $\Delta t=t_{i+1}-t_i$ and the following definitions apply.
\begin{equation}\label{eq_FVP_27}
\Delta U_{tf} := \int_{t_i}^{t_{i+1}}d U_{tf}, \quad \Delta \rho_f := \int_{t_i}^{t_{i+1}}d \rho_f, \quad \Delta Q_f := \int_{t_i}^{t_{i+1}}d Q_f
\end{equation}
We emphasize that the time step $\Delta t$ considered here is small, i.e., $\Delta t \ll 1$.

Denoting $\int_{t_i}^{t_{i+1}}d\left(\tilde{S}_c + \tilde{S}_k\right)$ by $\Sigma$, the path space average of $\Sigma$ may be found to be
\begin{equation}\label{eq_FVP_28}\begin{split}
 \langle\Sigma\rangle \approx &\frac{V\pi\nup\Delta t}{\theta_{ci}}\left(1+\frac{\sigma_{\theta_{ci}}^2}{\theta_{ci}^2}\right) -A\dot{\rho}_i\Delta t\left[\left(\frac{e_D}{\theta_{ci}}+\ln\left(a^2\rho_i\right)\right)+\frac{e_D\sigma_{\theta_{ci}^2}}{\theta_{ci}^3}-\frac{\sigma_{\rho_i}^2}{\rho_i^2}\right]\\ & + Q_i\Delta t \left[\left(\frac{1}{\theta_{k}}-\frac{1}{\theta_{ci}}\right)-\frac{\sigma_{\theta_{ci}}^2}{\theta_{ci}^2}\right],
\end{split}\end{equation}
where $\sigma_{\theta_{ci}}^2$ and $\sigma_{\rho_i}^2$ are respectively the variances of the effective temperature and dislocation density at time $t_i$. In deriving the above expression for $\langle\Sigma\rangle$, we have used an expression of the form
\begin{equation}\label{eq_FVP_29}\begin{split}
\langle f\left(X\right)g\left(Y\right)\rangle = & f\left(\bar{X}\right)g\left(\bar{Y}\right) +  f^{\prime}\left(\bar{X}\right) g^{\prime}\left(\bar{Y}\right)\text{Cov}\left(X,Y\right) \\&+ \frac{1}{2}f^{\prime \prime}\left(\bar{X}\right)g\left(\bar{Y}\right)\sigma_X^2 + \frac{1}{2}g^{\prime \prime}\left(\bar{Y}\right)f\left(\bar{X}\right)\sigma_Y^2,
\end{split}\end{equation}
where $f$ and $g$ are two sufficiently smooth functions and $X$, $Y$ are two random variables. We denote by $\bar{X}$ and $\bar{Y}$ the means of $X$ and $Y$, by $\sigma_X^2$ and $\sigma_Y^2$ their respective variances and with $\text{Cov}\left(X,Y\right)$ the covariance of $X$ and $Y$. In Eq. \eqref{eq_FVP_28}, terms containing $\sigma_{\theta_{ci}}^2$ and $\sigma_{\rho_i}^2$ are expected to be small in comparison with the others. We, thus, make use of the following approximation,
\begin{equation}\label{eq_FVP_30}\begin{split}
 \langle\Sigma\rangle \approx \frac{V\pi\nup\Delta t}{\theta_{ci}} -A\dot{\rho}_i\Delta t \left(\frac{e_D}{\theta_{ci}}+\ln\left(a^2\rho_i\right)\right) + Q_i\Delta t \left(\frac{1}{\theta_{k}}-\frac{1}{\theta_{ci}}\right)
\end{split}\end{equation}
Note that if we integrated the left hand side of Eq. \eqref{eq_FVP_16} over the small time interval $\left[t_i,t_{i+1}\right)$, the integral would be recovered as the path space average of entropy production, given in Eq. \eqref{eq_FVP_30}, i.e. the stochastic formulation obtains the deterministic entropy generation in the mean. However, the stochastic approach or the path space formulation provides us with even more information about the system via the higher order moments.

We compute, using Eq. \eqref{eq_FVP_29}, the variance of $\Sigma$, which, upon neglecting the terms of order $\left(\Delta t\right)^2$ and higher, takes the following form.
\begin{equation}\label{eq_FVP_31}\begin{split}
\text{Var}\left(\Sigma\right)=\Big\langle\Big(\Sigma - \langle\Sigma\rangle\Big)^2\Big\rangle \approx &\frac{1}{\theta_{k}^2}\sigma_{\Delta U_{tf}}^2 + A^2\left[\frac{e_D}{\theta_{k}}+\ln\left(a^2\rho_i\right)\right]^2\sigma_{\Delta \rho_f}^2 \\&+ \left(V C_{\text{eff}}\right)^2\left(\frac{1}{\theta_{k}}-\frac{1}{\theta_{ci}}\right)^2\sigma_{\Delta \theta_{cf}}^2
\end{split}\end{equation}
Here, $\sigma_{\Delta U_{tf}}^2$, $\sigma_{\Delta \rho_f}^2$ and $\sigma_{\Delta \theta_{cf}}^2$ are respectively the variances of $\Delta U_{tf}$, $\Delta \rho_f$ and $\Delta \theta_{cf}$. Higher moments of $\Sigma$ are of order more than $\left(\Delta t\right)^2$ and accordingly we neglect them.

Using Eq. \eqref{eq_FVP_30} and \eqref{eq_FVP_31}, the IFT, i.e, Eq. \eqref{eq_FVP_25}, may now be simplified to arrive at the following form.
\begin{equation}\label{eq_FVP_32}
\langle\Sigma\rangle \approx \frac{1}{2} \text{Var}\left(\Sigma\right)
\end{equation}
Eq. \eqref{eq_FVP_32}, in essence, expresses a fluctuation-dissipation type relation pertinent to the present formulation of metal viscoplasticity. We postulate that Eq. \eqref{eq_FVP_32} is satisfied in the following way.
\begin{equation}\label{eq_FVP_33}
\frac{V\pi\nup\Delta t}{\theta_{ci}} = \frac{1}{2\theta_{k}^2}\sigma_{\Delta U_{tf}}^2
\end{equation}
\begin{equation}\label{eq_FVP_34}
A\dot{\rho}_i\Delta t \left(\frac{e_D}{\theta_{ci}}+\ln\left(a^2\rho_i\right)\right) = -\frac{1}{2}A^2\left(\frac{e_D}{\theta_{k}}+\ln\left(a^2\rho_i\right)\right)^2\sigma_{\Delta \rho_f}^2
\end{equation}
and
\begin{equation}\label{eq_FVP_35}
Q_i\Delta t \left(\frac{1}{\theta_{k}}-\frac{1}{\theta_{ci}}\right) = \frac{1}{2}\left(V C_{\text{eff}}\right)^2\left(\frac{1}{\theta_{k}}-\frac{1}{\theta_{ci}}\right)^2\sigma_{\Delta \theta_{cf}}^2
\end{equation}

Eq. \eqref{eq_FVP_34} and \eqref{eq_FVP_35} lead respectively to equations for dislocation density evolution and heat flux between the subsystems as
\begin{equation}\label{eq_FVP_36}
\dot{\rho}_i=  -\frac{1}{2}A\left(\frac{e_D}{\theta_{k}}+\ln\left(a^2\rho_i\right)\right)^2 \left(\frac{e_D}{\theta_{ci}}+\ln\left(a^2\rho_i\right)\right)^{-1} \sigma_{\Delta \rho_f}^2 /\Delta t
\end{equation}
\begin{equation}\label{eq_FVP_37}
Q_i = \frac{1}{2}\left(V C_{\text{eff}}\right)^2\left(\frac{1}{\theta_{k}}-\frac{1}{\theta_{ci}}\right)\sigma_{\Delta \theta_{cf}}^2/\Delta t
\end{equation}
Using the approximation as carried out in Eq. \eqref{eq_FVP_17}, we rewrite Eq. \eqref{eq_FVP_36} as
\begin{equation}\label{eq_FVP_36_b}
 \dot{\rho}_i=  \frac{1}{2}A\left(1-\frac{\rho_i}{\rho_0\left(\theta_{k}\right)}\right)^2 \left(1-\frac{\rho_i}{\rho_0\left(\theta_{ci}\right)}\right)^{-1} \sigma_{\Delta \rho_f}^2 /\Delta t
\end{equation}
where, $\rho_0\left(\theta_{k}\right) = \frac{1}{a^2}\exp\left({-\frac{e_D}{\theta_{k}}}\right)$ and $\rho_0\left(\theta_{ci}\right) = \frac{1}{a^2}\exp\left({-\frac{e_D}{\theta_{ci}}}\right)$.

In the absence of an explicit expression for the path space distribution, we presently rely on the available experimental evidence and other conditions to find out, at least approximately, the unknown variances that appear in Eq. \eqref{eq_FVP_36} and \eqref{eq_FVP_37}. One important experimental data pertains to the initial hardening rate. For example in the case of oxygen-free high conductivity (OFHC) copper, in units of the shear modulus ($\mu$), it is roughly independent of both temperature and strain rate, and has a magnitude $0.05$, i.e.,
\begin{equation}\label{eq_FVP_38}
\frac{1}{\mu} \left(\frac{\dot\pi}{\nup}\right)_{\text{onset}}\approx 0.05
\end{equation}
For $\pi = \mu_T b\sqrt{\rho}\,\nu$ (see Eq. \eqref{eq59}), and at onset $\dot{\rho}_i=  A \sigma_{\Delta \rho_f}^2 /\left(2\Delta t\right)$, we get from Eq. \eqref{eq_FVP_38}
\begin{equation}\label{eq_FVP_39}
\sigma_{\Delta \rho_f}^2 \approx \frac{1}{5}\frac{\mu}{\left(\mu_T b\, \nu\right)^2}\frac{\pi\nup\Delta t}{A}
\end{equation}
It has been experimentally observed that the flow stress for some FCC metals (e.g. OFHC copper) increases dramatically when the strain rate exceeds a critical value. For copper, when the strain rate approaches $10^4/\text{S}$, the growth rate of dislocation generation begins to accelerate, leading to an abrupt increase of the dislocation density and a consequently sharp upturn of the flow stress. The expression for $\sigma_{\Delta \rho_f}^2 $ should therefore include this information too along with the initial hardening rate. In order to include the strain rate sensitivity, our proposal here is of the following form.
\begin{equation}\label{eq_FVP_39_b}
\sigma_{\Delta \rho_f}^2 \approx \frac{1}{5}\frac{\mu}{\left(\mu_T b\, \nu\right)^2}\frac{\pi\nup\Delta t}{A}\left(1+\frac{\nup}{\nup_{cr}}\right)
\end{equation}
$\nup_{cr}$ denotes the critical strain rate where a sharp upturn of the flow stress occurs.

To determine $\sigma_{\Delta \theta_{cf}}^2$, we make use of the steady state condition at which $\dot \theta_{c}$ and $\dot \rho$ is zero. Taking the path space average of Eq. \eqref{eq_FVP_21} and using Eq. \eqref{eq_FVP_37}, we get
\begin{equation}\label{eq_FVP_40}
 \sigma_{\Delta \theta_{cf}}^2 = \pi\nup \frac{2\theta_{c}^{ss}\theta_{k}}{C_\text{eff}\left(\theta_{c}^{ss}-\theta_{k}\right)}\Delta t,
\end{equation}
where $\theta_{c}^{ss}$ is the effective temperature in the steady state.

\section{Continuum formulation of viscoplasticity}\label{sec_Cont_form}
Extending the derivation of viscoplasticity given in section \ref{sec_Fluctuation_plast}, which considers an idealized homogeneous deformation and constant kinetic-vibrational temperature, we now set up a full fledged three dimensional formulation for viscoplastic deformation in FCC metals. This formulation accommodates a spatially inhomogeneous finite elasto-viscoplastic deformation, as well as an evolving kinetic-vibrational temperature.
\subsection{Kinematics}\label{kinematics}

Let ${\mathcal{B}_{0}}\subset {{\mathbb{R}}^{3}}$ be the reference configuration of a body at time ${{t}_{0}}$. Macroscopically, upon elasto-viscoplastic deformation at time $t$, each material point $\xb$ in this continuum is mapped to the point $\yb$ in its spatial configuration, $\mathcal{B}_t$. Specifically, the deformation map $\chib$, given by $\yb=\chib\left( \xb,t \right)$, is assumed to be smooth, one-one and onto. It therefore possesses a unique inverse. The deformation gradient ($\Fb$), velocity ($\vb$) and velocity gradient ($\Lb$) fields are defined as follows.
\begin{equation}\label{eq1}
\Fb := \nabla_{\xb}\chib  \quad\quad \vb := \dot{\chib} \quad\quad \Lb := \nabla_{\yb}\vb = \dot\Fb\Fb^{-1} ,
\end{equation}
where $\nabla_{\xb},\nabla_{\yb}$ and superposed dot respectively denote the gradient with respect to the material coordinate $\xb$, the gradient with respect to the spatial coordinate $\yb$ and the material time derivative.

In order to separate out the elastic and viscoplastic parts of $\Fb$, we opt for its multiplicative decomposition \citep{lee1969elastic}:
\begin{equation}\label{eq2}
\Fb=\Feb\Fpb ,
\end{equation}
where $\Fpb(\xb)$, a local plastic deformation, carries the material to a coherent structure residing in a relaxed intermediate configuration space and $\Feb(\xb)$ represents the subsequent rotation and stretching of the coherent structure. With the decomposition given in Eq. \eqref{eq2}, the velocity gradient $\Lb$ may be shown to admit the following decomposition.
\begin{equation}\label{eq3}
\Lb = \Leb + \Feb\Lpb\invFeb ,
\end{equation}
where $\Leb=\dot\Feb\invFeb$ and $\Lpb=\dot\Fpb\invFpb$. Elastic and viscoplastic rate of deformation tensors, ($\Deb$ and $\Dpb$), and spin tensors, ($\Web$ and $\Wpb$), are defined as:
\begin{equation}\label{eq4} \begin{split}
&\Deb = \frac{1}{2}\left(\Leb + {\Leb}^\text{T}\right) \quad \Dpb = \frac{1}{2}\left(\Lpb + {\Lpb}^\text{T}\right) \\\text{and} \quad
&\Web = \frac{1}{2}\left(\Leb - {\Leb}^\text{T}\right) \quad \Wpb = \frac{1}{2}\left(\Lpb - {\Lpb}^\text{T}\right)
\end{split}
\end{equation}

We adopt two kinematical assumptions concerning the plastic flow: \emph{incompressibility} and \emph{irrotationality}. These two assumptions translate to the following conditions.
\begin{equation}\label{eq5}
\det{\Fpb} =1 \quad\quad \text{tr}{\Lpb} = 0   \quad\quad \Wpb = \mathbf{0}
\end{equation}
Plastic irrotationality implies $\Lpb = \Dpb$ and consequently $\dot\Fpb = \Fpb\Dpb$. Defining a scalar $\nup$ and a tensor $\Npb$ as
\begin{equation}\label{eq6}
\nup := \left|\Dpb\right| \quad \text{and} \quad \Npb := \frac{\Dpb}{\nup}
\end{equation}
Eq. \eqref{eq3} may be recast as
\begin{equation}\label{eq7}
\nabla_\yb\vb = \Lb = \Leb + \nup\Feb\Npb\invFeb
\end{equation}

In line with the definition of Green-Lagrange strain tensor in elasticity, we define, for the present case, the elastic strain tensor $\Eeb$ as
\begin{equation}\label{eq8}
 \Eeb = \frac{1}{2}\left({\Feb}^\text{T}\Feb-\textbf{I}\right) = \frac{1}{2}\left(\Ceb-\textbf{I}\right)
\end{equation}

Since $\mathcal{B}_t$ evolves through an observable space, a change of the observer (i.e. a change of the reference frame $\mathcal{F}\rightarrow\mathcal{F}^*$) relates the observed spatial coordinates $\yb$ and $\yb^*$ as follows.
\begin{equation}\label{eq9}
\yb^*=\chib^*\left(\xb,t\right) = \Qb(t)\chib(\xb,t) + \mathbf{r}(t) = \Qb(t)\yb + \mathbf{r}(t) = \phi(\yb)
\end{equation}
Consequently, the tensorial quantities defined above transform as:
\begin{equation}\label{eq10}\begin{split}
&\Fb^* = \Qb\Fb, \quad  \dot{\Fb}^* = \dot{\Qb}\Fb + \Qb\dot\Fb, \quad \Lb^* = \dot\Qb\Qb^{\text{T}} + \Qb\Lb\Qb^{\text{T}}, \\
&{\Feb}^* = \Qb\Feb, \quad {\Fpb}^* = \Fpb, \quad {\Lpb}^* = \Lpb,  \quad {\Dpb}^* = \Dpb, \\
&{\Leb}^* = \dot\Qb\Qb^{\text{T}} + \Qb\Leb\Qb^{\text{T}}, \quad {\Deb}^* =  \Qb\Deb\Qb^{\text{T}},  \quad {\Web}^* = \dot\Qb\Qb^{\text{T}} + \Qb\Web\Qb^{\text{T}}
\end{split}
\end{equation}

\subsection{Equations of motion}\label{balance_law}
An essential ingredient in a continuum model is the equation describing the motion of the body. It is, generally, by Cauchy's hypothesis guaranteeing the existence of the traction vector, use of laws of balance for the linear and angular momenta and a subsequent localization argument that together lead to the equations of motion. However, following \citet{gurtin2010mechanics}, we take the balance of virtual power as a basic postulate in this work and derive the equations of motion thereon as presented below.

Let $\mathcal{P}_t$ denote an arbitrary subregion of $\mathcal{B}_t$ and $\partial\mathcal{P}_t$ its bounding surface with outward normal $\mathbf{n}$. The principle of virtual power is based on the balance of the external power input $\mathcal{W}_{\text{ext}}(\mathcal{P}_t)$ on $\mathcal{P}_t$ with the internal power generation $\mathcal{W}_{\text{int}}(\mathcal{P}_t)$ within $\mathcal{P}_t$. The rate quantities over which power is expended are $\vb$, $\Leb$ and $\nup$. However, these rates are not independent, rather constrained by Eq. \eqref{eq7}. Considering an elastic stress $\Seb$ and a scalar plastic stress $\pi$, the internal power expenditure within $\mathcal{P}_t$ is given by
\begin{equation}\label{eq11}
\mathcal{W}_{\text{int}}(\mathcal{P}_t) = \int_{\mathcal{P}_t}\left(\Seb:\Leb + J^{-1}\pi\nup\right)dv
\end{equation}
where $J=\det{\Fb} = \det{\Feb}$. $\pi$ and $\nup$ being quantities defined on the intermediate configuration, $J^{-1}$ is used to bring the plastic power expressed in per unit volume of the intermediate configuration to that of the current configuration.

Given the traction force vector $\mathbf{t}(\mathbf{n})$ working on the boundary, densities of the body force $\mathbf{b}_0$ and the inertial force $\rho_m\dot\vb$ ($\rho_m$ being the mass density) the external power input $\mathcal{W}_{\text{ext}}(\mathcal{P}_t)$ is given by the following.
\begin{equation}\label{eq12}
\mathcal{W}_{\text{ext}}(\mathcal{P}_t) =  \int_{\partial\mathcal{P}_t} \mathbf{t}(\mathbf{n})\cdot \vb ds + \int_{\mathcal{P}_t}\left(\mathbf{b}_0 - \rho_m\dot\vb \right)\cdot \vb dv
\end{equation}
We will henceforth denote $\left(\mathbf{b}_0 - \rho_m\dot\vb \right)$ by $\mathbf{b}$.

Considering virtual rate fields, $\tilde\vb$, $\tildeLeb$ and $\tildenup$, consistent with the the restriction
\begin{equation}\label{eq13}
 \nabla_\yb\tilde\vb  = \tildeLeb + \tildenup\Feb\Npb\invFeb
\end{equation}
the principle of virtual power balance for the region $\mathcal{P}_t$ is stated as
\begin{equation}\label{eq14}
 \mathcal{W}_{\text{ext}}(\mathcal{P}_t, \mathcal{V})= \mathcal{W}_{\text{int}}(\mathcal{P}_t, \mathcal{V}) \quad \forall \mathcal{V}
\end{equation}
$\mathcal{V}$ above is the list of virtual velocities: $\mathcal{V} = \left(\tilde\vb, \tildeLeb, \tildenup\right)$.

Other than the virtual power principle, the invariance of $\mathcal{W}_{\text{int}}(\mathcal{P}_t, \mathcal{V})$ under a change of the observer should also be satisfied. This leads to:
\begin{equation}\label{eq15}
 \mathcal{W}_{\text{int}}(\mathcal{P}_t, \mathcal{V}) = \mathcal{W}^*_{\text{int}}(\mathcal{P}^*_t, \mathcal{V}^*)
\end{equation}
Using the definition given in Eq. \eqref{eq11}, Eq. \eqref{eq15} is rewritten as
\begin{equation}\label{eq16}
\int_{\mathcal{P}_t}\left(\Seb:\tildeLeb + J^{-1}\pi\tildenup\right)dv(\yb) = \int_{\mathcal{P}^*_t}\left({\Seb}^*:\tilde{\Lb}^{e*} + {J^*}^{-1}\pi^*{\tilde\nu}^{p*}\right)dv(\yb^*)
\end{equation}
Under a change of the observer, $\pi\nup$ remains invariant. Therefore Eq. \eqref{eq16} takes the following form.
\begin{equation}\label{eq17}\begin{split}
 \int_{\mathcal{P}_t}\Seb(\yb):\tildeLeb(\yb) dv(\yb) &= \int_{\mathcal{P}^*_t}{\Seb}^*(\yb^*):\tilde{\Lb}^{e*}(\yb^*) dv(\yb^*) \\&= \int_{\mathcal{P}^*_t}{\Seb}^*(\yb^*):\left(\dot\Qb\Qb^{\text{T}} + \Qb\tildeLeb \left(\phi^{-1}(\yb^*)\right)\Qb^{\text{T}}\right) dv(\yb^*) \\&= \int_{\mathcal{P}_t}{\Seb}^*(\phi(\yb)):\left(\dot\Qb\Qb^{\text{T}} + \Qb\tildeLeb \left(\yb\right)\Qb^{\text{T}}\right) dv(\yb)
 \end{split}
\end{equation}
Equivalently one may write the following local form exploiting the arbitrariness of $\mathcal{P}_t$.
\begin{equation}\label{eq18}
 \Seb:\tildeLeb = {\Seb}^*:\left(\dot\Qb\Qb^{\text{T}} + \Qb\tildeLeb\Qb^{\text{T}} \right)  = \Qb^{\text{T}} {\Seb}^*\Qb:\tildeLeb + {\Seb}^*: \dot\Qb\Qb^{\text{T}}
\end{equation}
Without a loss of generality, the choice of a change of frame such that $\Qb$ is constant and the use of Eq. \eqref{eq18} yield
\begin{equation}\label{eq19}
 {\Seb}^* =  \Qb\Seb\Qb^{\text{T}}
\end{equation}
Substituting Eq. \eqref{eq19} in Eq. \eqref{eq18}, we get ${\Seb}^*: \dot\Qb\Qb^{\text{T}} = 0$. For arbitrary $\Qb$, $\dot\Qb\Qb^{\text{T}}$ is an arbitrary skew tensor and thus we conclude that ${\Seb}^* = {\Seb}^{*\text{T}} $ and hence, using Eq. \eqref{eq19}, that
\begin{equation}\label{eq20}
 {\Seb} = {\Seb}^{\text{T}}
\end{equation}
\subsubsection{Macroscopic force balance}
To derive the macroscopic force balance, consider the virtual velocity  associated with the viscoplastic deformation to be zero, i.e. $\tildenup=0$. Following Eq. \eqref{eq13}, the admissible virtual elastic distortion rate $\tildeLeb$ is given by
\begin{equation}\label{eq21}
\nabla_{\yb}\tilde\vb=\tildeLeb,
\end{equation}
where $\tilde\vb$ is arbitrary. Virtual power balance (see Eq. \eqref{eq14}), in this case, takes the following form.
\begin{equation}\label{eq22}
\int_{\partial\mathcal{P}_t} \mathbf{t}(\mathbf{n})\cdot \tilde\vb ds + \int_{\mathcal{P}_t}\mathbf{b}\cdot \tilde\vb dv = \int_{\mathcal{P}_t}\Seb: \nabla_{\yb}\tilde\vb dv
\end{equation}
Using divergence theorem on the integral appearing on the right hand side of Eq. \eqref{eq22}, the equation may be recast as
\begin{equation}\label{eq23}
\int_{\partial\mathcal{P}_t} \left(\mathbf{t}(\mathbf{n})- \Seb\mathbf{n}\right)\cdot \tilde\vb ds + \int_{\mathcal{P}_t}\left(\nabla_\yb\cdot \Seb + \mathbf{b}\right)\cdot \tilde\vb dv = \mathbf{0}
\end{equation}
Upon localization based on the arbitrariness of $\mathcal{P}_t$ , one is led to the macroscopic force balance and traction-stress relation as follows.
\begin{equation}\label{eq24}
 \nabla_\yb\cdot \Seb + \mathbf{b} = \mathbf{0}
\end{equation}
\begin{equation}\label{eq25}
\Seb\mathbf{n} = \mathbf{t}(\mathbf{n})
\end{equation}
Note that the arbitrariness of the virtual velocity $\tilde\vb$ is also used to write Eq. \eqref{eq24} and \eqref{eq25}. The elastic stress $\Seb$ satisfying Eq. \eqref{eq24} and \eqref{eq25} along with the properties given in Eq. \eqref{eq19} and \eqref{eq20} may be identified with the classical Cauchy stress $\Tb$ and therefore, in what follows, we will replace $\Seb$ by $\Tb$.
\subsubsection{Microscopic force balance}
The microscopic force balance may be derived considering $\tilde\vb=\mathbf{0}$. Eq. \eqref{eq13} and the virtual power balance Eq. \eqref{eq14}, in this case, take the following forms.
\begin{equation}\label{eq26}
 \tildeLeb =- \tildenup\Feb\Npb\invFeb
 \end{equation}
 \begin{equation}\label{eq27}
 \int_{\mathcal{P}_t}\left(\Tb:\tildeLeb + J^{-1}\pi\tildenup\right) dv = 0
\end{equation}
Arbitrariness of $\mathcal{P}_t$ again allows localization of Eq. \eqref{eq27}, which together with Eq. \eqref{eq26}, takes the following form.
\begin{equation}\label{eq28}\begin{split}
J^{-1}\pi\tildenup = - \Tb:\tildeLeb =  - \Tb:\left(- \tildenup\Feb\Npb\invFeb\right) &= \left({\Feb}^{\text{T}}\Tb{\Feb}^{\text{-T}}:\Npb\right)\tildenup \\ &= \left({\Feb}^{\text{T}}\Tb_0{\Feb}^{\text{-T}}:\Npb\right)\tildenup
\end{split}\end{equation}
The last equality in Eq. \eqref{eq28} follows from the fact that $\text{tr}\Npb = 0$. In this last term, $\Tb_0$ is the deviatoric part of $\Tb$. $\tildenup$ being arbitrary, Eq. \eqref{eq28} leads to the following microscopic force balance.
\begin{equation}\label{eq29}
\pi = J{\Feb}^{\text{T}}\Tb_0{\Feb}^{\text{-T}}:\Npb
\end{equation}

Apart from the Cauchy stress $\Tb$, one may also define a second Piola-Kirchoff type stess $\Teb$ and Mendel stress $\mathbf{M}^e$ as
\begin{equation}\label{eq30}
\Teb := J{\Feb}^{-1}\Tb{\Feb}^{\text{-T}} \quad \text{and} \quad \mathbf{M}^e :=  \Ceb\Teb
\end{equation}
In terms of the Mendel stress, the microscopic force balance Eq. \eqref{eq29} assumes the form:
\begin{equation}\label{eq31}
\pi = \mathbf{M}^e_0:\Npb
\end{equation}
Here $\mathbf{M}^e_0$ is the deviatoric part of $\mathbf{M}^e$.

\subsection{Constitutive relations}\label{constitution}
The constitutive relation plays a pivotal role in continuum material modeling. It provides closure to the equations of motion by describing a stress-strain relationship for the given material. It is through the constitutive model that material behaviour, e.g. viscoplasticity in the present case, enters the continuum formulation. Herein we base our constitutive theory of metal viscoplasticity on the dislocation dynamics described through a single internal variable-- the dislocation density $\rho$. Instead of distinguishing the different types of dislocations, we will consider $\rho$ to include their effects in an averaged sense. The goal of constitutive modelling here is to express $\Tb$ and $\pi$ in terms of the kinematic quantities and material parameters, and also to find out the evolution equation for the dislocation density and the heat flux-temperature relation.

\subsubsection{First and second laws of thermodynamics}
As noted in the introduction, we take recourse to two-temperature thermodynamics to model the viscoplastic deformation in metals. To start with, we state the first law of thermodynamics or the internal energy balance as
\begin{equation}\label{eq32}
\rho_m\dot e = p_{\text{int}} - \nabla_\yb\cdot \mathbf{q},
\end{equation}
where $e$ , $p_{\text{int}}$ and $\mathbf{q}$ are respectively the specific internal energy, the internal power density and the heat flux vector. We have neglected the external heat source in writing out the first law. The internal power density is given by
\begin{equation}\label{eq33}
 p_{\text{int}} := \Tb:\Leb + J^{-1}\pi\nup = \Tb:\Deb + J^{-1}\pi\nup = J^{-1}\Teb:\dot{\mathbf{E}}^e + J^{-1}\pi\nup
\end{equation}
As opposed to the global description of first law (see Eq. \eqref{eq_FVP_14}) used in section \ref{sec_Fluctuation_plast}, Eq. \eqref{eq32} is a local representation of the same and is useful to describe spatial variations of fields.

Recalling that the total internal energy density $e$ and the heat flux $\mathbf{q}$ have contributions from two distinct sources, viz. the configurational and kinetic-vibrational subsystems, we consider the following additive decompositions.
\begin{equation}\label{eq34}
 e = e_c+e_k \quad \text{and} \quad \mathbf{q} = \mathbf{q}_c + \mathbf{q}_k,
\end{equation}
The subscripts $c$ and $k$ respectively denote the configurational and kinetic-vibrational contributions. Following \citet{kamrin2014two}, we split the energy balance equation \eqref{eq32} for the two subsystems as
\begin{equation}\label{eq35}
 \rho_m\dot e_c = J^{-1}\Teb:\dot{\mathbf{E}}^e + J^{-1}\pi\nup - q_{ck} - \nabla_\yb\cdot \mathbf{q}_c,
\end{equation}
and
\begin{equation}\label{eq36}
 \rho_m\dot e_k = q_{ck} - \nabla_\yb\cdot \mathbf{q}_k,
\end{equation}
When driven out of equilibrium by an external load, the scalar heat flux that gets generated from the configurational to the kinetic-vibrational subsystem is designated as  $q_{ck}$. $q_{ck}$ in fact acts as a source of heat for the kinetic-vibrational subsystem. It is readily seen that, when added, Eq. \eqref{eq35} and \eqref{eq36} recover the energy balance for the entire system (see Eq. \eqref{eq32}).

The constitutive relations must be compatible with the second law of thermodynamics too. With the hypothesis of local equilibrium, the second law inequality, in the absence of an external entropy source, may be written as
\begin{equation}\label{eq37}
\rho_m\dot\eta + \nabla_\yb\cdot\mathbf{j} \ge 0
\end{equation}
where $\eta$  is the specific entropy and $\mathbf{j}$ the entropy flux. Once more we adopt additive decompositions: $\eta = \eta_c+\eta_k$ and $\mathbf{j}=\mathbf{j}_c+\mathbf{j}_k$. With the decomposed entropy fluxes given by the following relations
\begin{equation}\label{eq38}
 \mathbf{j}_c = \frac{\mathbf{q}_c}{\theta_c} \quad \text{and} \quad \mathbf{j}_k = \frac{\mathbf{q}_k}{\theta_k},
\end{equation}
where $\theta_c$ and $\theta_k$ respectively denote the configurational (effective) and the kinetic-vibrational (ordinary) temperatures, we may write the entropy inequality as follows.
\begin{equation}\label{eq39}
 \rho_m\dot\eta_c +\rho_m\dot\eta_k + \nabla_\yb\cdot\left(\frac{\mathbf{q}_c}{\theta_c}\right) + \nabla_\yb\cdot\left(\frac{\mathbf{q}_k}{\theta_k}\right) \ge 0
\end{equation}
Substituting for $\nabla_\yb\cdot \mathbf{q}_c$ and $\nabla_\yb\cdot \mathbf{q}_k$ form Eq. \eqref{eq35} and \eqref{eq36}, Eq. \eqref{eq39} is recast as
\begin{equation}\label{eq40}\begin{split}
 \rho_m\dot\eta_c +\rho_m\dot\eta_k &+ \frac{1}{\theta_c}\left[J^{-1}\left(\Teb:\dot{\mathbf{E}}^e+\pi\nup\right)-q_{ck}-\rho_m\dot e_c\right] + \frac{1}{\theta_k}\left(q_{ck}-\rho_m\dot e_k\right) \\&-\frac{1}{\theta_c^2}\mathbf{q}_c\cdot\nabla_\yb\theta_c-\frac{1}{\theta_k^2}\mathbf{q}_k\cdot\nabla_\yb\theta_k \ge 0
 \end{split}
\end{equation}
Eq. \eqref{eq40} is a spatial description of the entropy inequality. In transforming this inequality to the intermediate frame representation, we denote $\rho_m^I = J\rho_m$ as the mass density of that configuration and
also define quantities relative to volumes, areas and lengths in this intermediate space as
\begin{equation}\label{41}\begin{split}
&e_c^I=\rho_m^I e_c, \quad e_k^I=\rho_m^I e_k, \quad \eta_c^I=\rho_m^I \eta_c, \quad \eta_k^I=\rho_m^I \eta_k, \quad q_{ck}^I = Jq_{ck} \\
&\mathbf{q}_c^I = J\invFeb\mathbf{q}_c, \quad \mathbf{q}_k^I = J\invFeb\mathbf{q}_k, \quad \mathbf{g}_c^I = \Fb^{e\text{T}}\nabla_\yb\theta_c, \quad \mathbf{g}_k^I = \Fb^{e\text{T}}\nabla_\yb\theta_k
\end{split}
\end{equation}
For ease of exposition, we drop the superscript $I$ from the notations, and write the intermediate space description of entropy imbalance as:
\begin{equation}\label{eq42}\begin{split}
 \dot\eta_c + \dot\eta_k &+ \frac{1}{\theta_c}\left(\Teb:\dot{\mathbf{E}}^e+\pi\nup-q_{ck}-\dot e_c\right) + \frac{1}{\theta_k}\left(q_{ck}-\dot e_k\right) \\&-\frac{1}{\theta_c^2}\mathbf{q}_c\cdot\mathbf{g}_c-\frac{1}{\theta_k^2}\mathbf{q}_k\cdot\mathbf{g}_k \ge 0
 \end{split}
\end{equation}
Intermediate space representations of Eq. \eqref{eq35} and \eqref{eq36} take the following forms.
\begin{equation}\label{eq43}
\dot e_c = \Teb:\dot{\mathbf{E}}^e + \pi\nup - q_{ck} - \nabla\cdot \mathbf{q}_c,
\end{equation}
\begin{equation}\label{eq44}
\dot e_k = q_{ck} - \nabla\cdot \mathbf{q}_k,
\end{equation}
Here $\nabla\cdot$ is the divergence operator defined on the intermediate space and its relation with spatial divergence is given by $\nabla\cdot\mathbf{u} = J\nabla_\yb\cdot\left(J^{-1}\Feb \mathbf{u}\right)$ for any vector field $\mathbf{u}$ defined on the intermediate space.

\subsubsection{Constitutive relations and evolution equations }
We take $e_c$ ,$e_k$, $\Eeb$ along with the dislocation density $\rho$ as independent variables describing the state of the viscoplastically deforming metal. We also consider the following dependencies
\begin{equation}\label{eq45}
\eta_c = \hat{\eta}_c\left(e_c, \Eeb, \rho\right), \quad \eta_k = \hat{\eta}_k\left(e_k\right),\quad \Teb=\hat{\mathbf{T}}^e\left(\Eeb\right),
\end{equation}
This allows us to expand Eq. \eqref{eq42} as follows.
\begin{equation}\label{eq46}\begin{split}
 &\left(\frac{\partial\hat{\eta}_c}{\partial e_c}-\frac{1}{\theta_c}\right)\dot e_c  + \left(\frac{\partial\hat{\eta}_k}{\partial e_k}-\frac{1}{\theta_k}\right)\dot e_k + \left(\frac{\partial\hat{\eta}_c}{\partial \Eeb}+\frac{1}{\theta_c}\Teb\right): \dot{\mathbf{E}}^e + \frac{\partial\hat{\eta}_c}{\partial \rho}\dot\rho \\ &+ \frac{1}{\theta_c}\pi\nup + \left(\frac{1}{\theta_k}-\frac{1}{\theta_c}\right)q_{ck} - \frac{1}{\theta_c^2}\mathbf{q}_c\cdot\mathbf{g}_c - \frac{1}{\theta_k^2}\mathbf{q}_k\cdot\mathbf{g}_k \ge 0
\end{split}
\end{equation}
The first three terms of the inequality \eqref{eq46} being linear in the independent rate quantities $\dot e_c$, $\dot e_k$ and $\dot{\mathbf{E}}^e$, the standard Coleman-Noll type argument \citep{coleman1963thermodynamics} ensuring non-negativity leads to the following.
\begin{equation}\label{eq47}
 \frac{1}{\theta_c} = \frac{\partial\hat{\eta}_c}{\partial e_c}, \quad\quad \frac{1}{\theta_k} = \frac{\partial\hat{\eta}_k}{\partial e_k},
\end{equation}
and
\begin{equation}\label{eq48}
Sym\left(\frac{\partial\hat{\eta}_c}{\partial \Eeb}+\frac{1}{\theta_c}\Teb\right) = \mathbf{0}
\end{equation}
By $Sym(\mathbf{A})$ we denote the symmetric part of the tensor $\mathbf{A}$. However, $\Eeb$ and $\Teb$ both being symmetric tensors, $\left(\frac{\partial\hat{\eta}_c}{\partial \Eeb}+\frac{1}{\theta_c}\Teb\right)$ itself is symmetric and thus Eq. \eqref{eq48} leads to:
\begin{equation}\label{eq49}
\Teb = -\theta_c\frac{\partial\hat{\eta}_c}{\partial \Eeb}
\end{equation}
Substituting Eq. \eqref{eq47} and \eqref{eq49} into \eqref{eq46}, the entropy imbalance reduces to
\begin{equation}\label{eq50}
 \frac{\partial\hat{\eta}_c}{\partial \rho}\dot\rho + \frac{1}{\theta_c}\pi\nup + \left(\frac{1}{\theta_k}-\frac{1}{\theta_c}\right)q_{ck} - \frac{1}{\theta_c^2}\mathbf{q}_c\cdot\mathbf{g}_c - \frac{1}{\theta_k^2}\mathbf{q}_k\cdot\mathbf{g}_k \ge 0.
\end{equation}
We now assume that $\hat{\eta}_c$ admits the following decomposition.
\begin{equation}\label{eq51}
 \hat{\eta}_c\left(e_c, \Eeb, \rho\right) = \hat{\eta}_\rho\left(\rho\right) + \hat{\eta}_\text{surr}\left(e_\text{surr}\right) \quad \text{with} \quad e_\text{surr}:= e_c- \hat{e}_\rho(\rho)-\hat{e}_e(\Eeb)
\end{equation}
The configurational entropy density and energy density associated with the dislocations, i.e.  $\hat{\eta}_\rho\left(\rho\right)$ and $\hat{e}_\rho(\rho)$, may be given by the following.
\begin{equation}\label{eq52}
 \hat{\eta}_\rho\left(\rho\right) = \frac{\rho}{L}-\frac{\rho}{L}\ln\left(a^2\rho\right)  \quad \text{and} \quad \hat{e}_\rho(\rho) = \frac{e_D\rho}{L}
\end{equation}
These expressions are obtained by dividing Eq. \eqref{eq_FVP_13} and \eqref{eq_FVP_10} by the volume $AL$. The elastic energy $\hat{e}_e(\Eeb)$ may be taken to be
\begin{equation}\label{eq53}
 \hat{e}_e(\Eeb) = \frac{1}{2}\lambda\left(\text{tr}\left(\Eeb\right)\right)^2 + \mu \Eeb:\Eeb,
\end{equation}
where $\lambda$ and $\mu$ are the Lam\'e parameters. This along with Eq. \eqref{eq49} and \eqref{eq51} leads to
\begin{equation}\label{eq53b}
\Teb = \lambda \text{tr}\left(\Eeb\right) \mathbf{I} + \mu \Eeb
\end{equation}
Using Eq. \eqref{eq51}- \eqref{eq52} and the fact that $\frac{\partial \hat{\eta}_\text{surr}}{\partial e_\text{surr}} = \frac{1}{\theta_c}$,  Eq.  \eqref{eq50} may be recast as
\begin{equation}\label{eq54}
 -\frac{1}{L}\left(\frac{e_D}{\theta_c}+\ln\left(a^2\rho\right)\right)\dot\rho + \frac{1}{\theta_c}\pi\nup + \left(\frac{1}{\theta_k}-\frac{1}{\theta_c}\right)q_{ck} - \frac{1}{\theta_c^2}\mathbf{q}_c\cdot\mathbf{g}_c - \frac{1}{\theta_k^2}\mathbf{q}_k\cdot\mathbf{g}_k \ge 0.
\end{equation}

\emph{Heat flux-temperature relations}:
To ensure non-negative entropy production, we demand that all the five terms appearing on the left hand side of \eqref{eq54} be individually non-negative. Choice of the constitutive relations for the heat fluxes in the form a Fourier-type law, as given in Eq. \eqref{eq55}, insures against negativity of the last three terms.
\begin{equation}\label{eq55}\begin{split}
&\mathbf{q}_c =  -k_c\mathbf{g}_c, \quad k_c = \hat{k}_c\left(\theta_c, \nup\right)\ge 0 \\
&\mathbf{q}_k =  -k_k\mathbf{g}_k, \quad k_k = \hat{k}_c\left(\theta_k\right)\ge 0 \\
&q_{ck} =  k_{ck}\left(\theta_c-\theta_k\right), \quad k_{ck} = \hat{k}_{ck}\left(\theta_c,\theta_k, \nup\right)\ge 0
\end{split}
\end{equation}
Using Eq. \eqref{eq_FVP_37} and \eqref{eq_FVP_40}, we may write $k_{ck} = \frac{\theta_c^{ss}}{\theta_c}\frac{\pi\nup}{\left(\theta_c^{ss}-\theta_k\right)}$.

\emph{Dislocation density evolution}:
With assured non-negativity of the last three terms in \eqref{eq54}, it only remains to do the same for the first two terms. Using Eq. \eqref{eq_FVP_36}, we may obtain the evolution of the dislocation density as
\begin{equation}\label{eq56}
\dot{\rho}=  -k_p\frac{\pi\nup}{\gamma_D} \left(1+\frac{\nup}{\nup_{cr}}\right)\left(\frac{e_D}{\theta_{k}}+\ln\left(a^2\rho\right)\right)^2 \left(\frac{e_D}{\theta_{c}}+\ln\left(a^2\rho\right)\right)^{-1}
\end{equation}
which guarantees non-negativity of the first term. Here $k_p$ is a non-negative number, and $\gamma_D = e_D/L$ the energy of dislocation per unit length. For copper, we get, using \eqref{eq_FVP_39_b}, $k_p = \frac{1}{10}\frac{\mu}{\left(\mu_T b\, \nu\right)^2}\gamma_D$. As in Eq. \eqref{eq_FVP_36_b}, we use, in lieu of Eq. \eqref{eq56}, the following approximate form to describe the dislocation density evolution.
\begin{equation}\label{eq56b}
\dot{\rho}=  -k_p\frac{\pi\nup}{\gamma_D} \left(1+\frac{\nup}{\nup_{cr}}\right)\left(1-\frac{\rho}{\rho_0\left(\theta_{k}\right)}\right)^2 \left(1-\frac{\rho}{\rho_0\left(\theta_{c}\right)}\right)^{-1}
\end{equation}

\emph{Constitutive relation for $\pi$}:
It is now remaining to ensure $\frac{1}{\theta_c}\pi\nup\ge 0$. By definition (see Eq. \eqref{eq6}), $\nup$ is non-negative and this is true for $\theta_c$ too. Therefore we are only required to frame the constitutive relation for $\pi$ such that the latter is always non-negative. To arrive at this constitutive relation for $\pi$, we take recourse to the theory of thermal activation of dislocations. For the present work, we only focus on FCC metals. The kinetic equation relating $\nup$ and $\pi$ for FCC metals may be given in the following Arrhenius form.
\begin{equation}\label{eq58}
\nup = \frac{\rho b l^*}{\tau_0}\exp\left[-\frac{k_BT_p}{\theta_k}\exp\left(-\frac{\pi}{\pi_T}\right)\right]
\end{equation}
Here $l^*$ is some mean free path or average distance between dislocations, $b$ the magnitude of the Burgers vector, $\tau_0^{-1}$ a microscopic attempt frequency of the order of $10^{12}$ per second. $k_BT_p$ is the thermal barrier of the potential in which the dislocation is trapped in absence of external driving. $k_B$ refers to the Boltzmann constant. $\pi_T$ is the Taylor stress given by $\pi_T = \mu_Tb\sqrt{\rho}$, where $\mu_T$ is proportional to the shear modulus $\mu$. For further details, see \citet{langer2010}. Inverting Eq. \eqref{eq58}, we get the constitutive relation for $\pi$ as
\begin{equation}\label{eq59}
 \pi = \mu_Tb\sqrt{\rho}\left[\ln\left(\frac{k_BT_p}{\theta_k}\right)-\ln\left\{\frac{1}{2}\ln\left(\frac{\rho b^2}{\tau_0^2{\nup}^2}\right)\right\}\right] = \mu_Tb\sqrt{\rho}\,\nu,
\end{equation}
The term within the square brackets, denoted by $\nu$, expresses the temperature and strain rate dependence of $\pi$. $\nu > 0$, which is the case in general, ensures the nonnegativity of $\pi$.

\emph{Temperature evolution}:
From Eq. \eqref{eq43} and \eqref{eq44}, evolution equations for the effective and kinetic-vibrational temperatures may be derived. In doing so, we assume dependencies $e_k = \hat{e}_k\left(\theta_k\right)$ and $e_c = \hat{e}_c\left(\theta_c, \Eeb, \rho\right) $. To be more precise, $e_c$ also depends on $\theta_k$ as the material parameters, e.g. $\lambda$, $\mu$, $e_D$ etc., depend on $\theta_k$. This allows to compute the energy rates as follows.
\begin{equation}\label{eq60}
\dot e_k= \frac{\partial\hat{e}_k}{\partial\theta_k}\dot{\theta}_k = \hat{c}_k\left(\theta_k\right)\dot{\theta}_k
\end{equation}
\begin{equation}\label{eq61}
\dot e_c= \frac{\partial \hat{e}_c}{\partial \theta_c}\dot{\theta}_c + \frac{\partial \hat{e}_c}{\partial \Eeb}:\dot{\mathbf{E}}^e + \frac{\partial \hat{e}_c}{\partial \rho}\dot{\rho} +  \frac{\partial \hat{e}_c}{\partial \theta_k}\dot{\theta}_k
\end{equation}
Taking the partial derivative of Eq. \eqref{eq51} with respect to $\Eeb$, we get
\begin{equation}\label{eq62}
\frac{\partial \hat{\eta}_c}{\partial \Eeb} = \frac{\partial \hat{\eta}_{\text{surr}}}{\partial e_{\text{surr}}}\left(-\frac{\partial \hat{e}_c\left(\Eeb\right)}{\partial \Eeb}\right) = - \frac{1}{\theta_c}\frac{\partial \hat{e}_c}{\partial \Eeb}
\end{equation}
We rewrite Eq. \eqref{eq61} using Eq. \eqref{eq49}, \eqref{eq52} and \eqref{eq62} as
\begin{equation}\label{eq63}
\dot e_c= \hat{c}_c\dot{\theta}_c + \Teb:\dot{\mathbf{E}}^e + \gamma_D\dot{\rho} + \frac{\partial \hat{e}_c}{\partial \theta_k}\dot{\theta}_k
\end{equation}
In Eq. \eqref{eq60} and \eqref{eq63}, $\hat{c}_k$ and $\hat{c}_c$ respectively denote the specific heat of the kinetic-vibrational subsystem and the configurational subsystem. Using Eq. \eqref{eq60}, \eqref{eq63} and \eqref{eq55}, the energy balance equations given by Eq. \eqref{eq43} and \eqref{eq44} may be written in the form of temperature evolution equations as
\begin{equation}\label{eq64}
\hat{c}_c\dot{\theta}_c = \pi\nup - \gamma_D\dot\rho -k_{ck}\left(\theta_c - \theta_k\right) +  \frac{\partial \hat{e}_c}{\partial \theta_k}\dot{\theta}_k  +\nabla\cdot\left(k_c\mathbf{g}_c\right)
\end{equation}
and
\begin{equation}\label{eq65}
\hat{c}_k\dot{\theta}_k = k_{ck}\left(\theta_c - \theta_k\right) + \nabla\cdot\left(k_k\mathbf{g}_k\right)
\end{equation}

\section{Extension to BCC metals and multiple dislocation densities}\label{sec_extension}
\subsection{BCC metals}
The continuum viscoplasticity model discussed thus far is restricted to FCC metals alone. Specifically, it is the way the constitutive model for $\pi$ is written that results in this restriction. We briefly suggest the required modifications of the constitutive relation for $\pi$ so as to render the present model applicable to BCC metals. While, for FCC metals, $\pi$ is basically of thermal origin, it has both athermal and thermal components for BCC metals. The athermal portion (denoted by $\pi_{\text{ath}}$) originates form a combination of dislocation-dislocation interaction, grain/sub-grain size effect and twining and may be expressed as
\begin{equation}\label{eq_BCC_1}
\pi_{\text{ath}} = \alpha\mu b\sqrt\rho,
\end{equation}
where $\alpha$ is a generalized interaction coefficient dependent on dislocation density, grain and sub-grain size, mean distance between twins and temperature.
The thermal portion (denoted by $\pi_{\text{th}}$) is directly related to the plastic strain rate and temperature sensitivity and a kinetic relation as given below (similar to Eq. \eqref{eq58}) holds.
\begin{equation}\label{eq_BCC_2}
\nup = \frac{\rho b l^*}{\tau_0}\exp\left[-\frac{k_BT_p}{\theta_k}\exp\left(-\frac{\pi_{\text{th}}}{\pi_p}\right)\right]
\end{equation}
Here $\pi_p$ is the Peierls stress proportional to $\mu$. Inverting Eq. \eqref{eq_BCC_2}, $\pi_{\text{th}}$ is obtained and the microscopic stress is given by $\pi = \pi_{\text{ath}} + \pi_{\text{th}}$.
\subsection{Multiple dislocation densities}
 Using a single dislocation density may prove inadequate in deriving predictive models that are meant to work over a wide range of strain rates and temperatures. The way a dislocation-type predominates in a specific situation and contributes to the hardening phenomena demands identifying the mobile and immobile parts, and model their interaction and evolution separately. Towards this, we may try to generalize the present theory to accommodate a two-internal variable representation for the configuration, viz. the mobile dislocation density $\rho_m$ and the sessile dislocation density $\rho_s$. This extension entails a systematic modification of the present one-parameter proposal, which we will undertake separately. Nevertheless, an important modification, perhaps the most so, is the definition of entropy associated with the dislocations (see Eq. \eqref{eq_FVP_13}). We suggest it to be as follows.
\begin{equation}\label{eq_mult_1}\begin{split}
S_0 = \ln\left(\Omega\right)=\ln\frac{N_{ls}!}{N_{md}!N_{sd}!\left(N_{ls}-N_{md}-N_{sd}\right)!} \approx \ln\frac{N_{ls}^{N_d}}{N_{md}!N_{sd}!},
\end{split}\end{equation}
The approximation follows from the fact that the total number of dislocations $N_d = N_{md} + N_{sd}$ is much less than that of lattice sites $N_{ls}$. By $N_{md} $ and $ N_{sd}$ we denote the number of mobile and sessile dislocations respectively. Applying Stirling's approximation, Eq. \eqref{eq_mult_1} is rewritten, in terms of the densities, as
\begin{equation}\label{eq_mult_2}
 S_0 = -A\rho_m\ln\left(a^2\rho_m\right) -A\rho_s\ln\left(a^2\rho_s\right) + A\left(\rho_m+\rho_s\right)
 \end{equation}

\section{Numerical Illustration}\label{sec_numerical}
The primary aim of the article has been on developing a viscoplasticity model for metals, consistent with the fluctuation relation, leaving a detailed numerical work with the model for a separate study. For mere demonstration withal, we present an elementary numerical simulation-- a copper cube undergoing simple shear deformation. The time dependent deformation map is given by
\begin{equation}\label{eq_num_1}
x = X + \alpha t Y, \quad y = Y, \quad z = Z,
\end{equation}
where $\left(X,Y,Z\right)$ are the coordinates of the reference configuration at time $t=0$ and they gets deformed to $\left(x,y,z\right)$ at $t$. $\alpha$ is a constant that governs the strain rate. Deformation given by Eq. \eqref{eq_num_1} leads to homogeneous fields for deformation, strain, stress etc. and a constant velocity field. In the absence of body force, the macroscopic force balance, i.e. Eq. \eqref{eq24}, therefore gets trivially satisfied. Homogeneous deformation condition leads to zero temperature gradient in both configurational and kinetic-vibrational subsystems and consequently the divergence terms (heat flux) that appear in Eq. \eqref{eq64} and \eqref{eq65} vanish. We solve a set of nonlinear ordinary differential equations (Eq. \eqref{eq56b}, Eq. \eqref{eq64} and \eqref{eq65} with vanishing divergence terms), along with the the microscopic force balance (Eq. \eqref{eq31}) and constitutive relations (Eq. \eqref{eq53b}, \eqref{eq59}) to obtain the quantitative description of elasto-viscoplastic deformation of the cube undergoing simple shear.

  \begin{figure}[!htp]
    \begin{center}
            \includegraphics[width=0.7\textwidth]{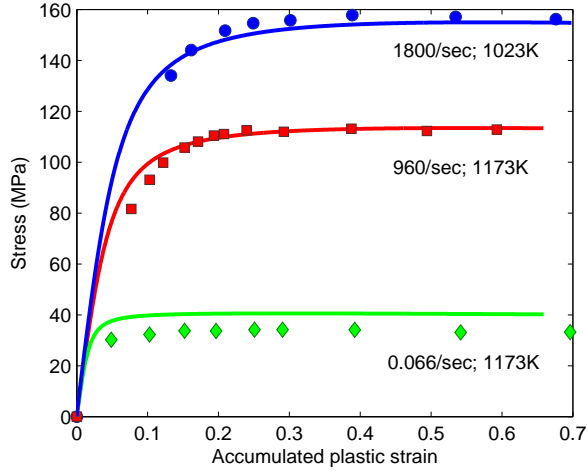}
    \end{center}
    \caption{Stress vs. accumulated plastic strain at high temperature}%
     \label{stress_strain_high_temp}
   \end{figure}

\begin{figure}[!htp]
\begin{center}
\includegraphics[width=0.7\textwidth]{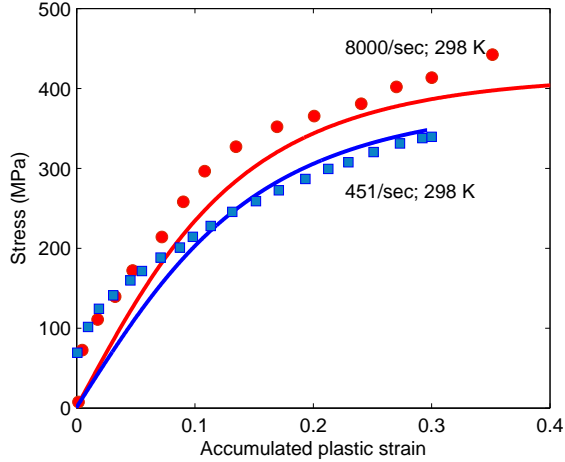}
\end{center}
\caption{Stress vs. accumulated plastic strain at room temperature}%
\label{stress_strain_room_temp}
\end{figure}

We present the evolution of $\pi$ with plastic deformation in Fig. \ref{stress_strain_high_temp} and Fig. \ref{stress_strain_room_temp}. Plastic deformation is quantified as accumulated plastic strain obtained upon integrating $\nup$. Fig. \ref{stress_strain_high_temp} shows the stress-strain plot at high temperature, whereas Fig. \ref{stress_strain_room_temp} presents the low temperature response. Both the figures include plots of plastic deformation at considerably varied strain rates. The following parameter values and relations are used for the simulations. Poisson's ratio 0.355, $T_P=40800$ K, $(a/b)\tau_0 = 10^{-12}$ sec, $\gamma_D = \mu b^{\prime 2}$, $\mu \approx \left(a/b^\prime\right)\bar{\mu}_T$, $\bar{\mu}_T = (b/a)\mu_T$, $a/b^\prime = 32$ at 298 K, and 31 at 1023 K and 1173 K,  $\bar{\mu}_T = 1500$ MPa at 298 K,  1460 MPa at 1023 K and 1350 MPa at 1173 K. For the range of temperature considered, $e_D$ is taken as $3.9\times10^{-23}(T_k)^{1.05}$ J, where $T_k = \theta_k/k_B$. Specific heat of the kinetic vibrational system  0.39 KJ/Kg K. $a^2 \rho \left(t=0\right) = 10^{-7}$, $\theta_c \left(t=0\right)/ e_{Dm} = 0.25$ and $\theta_c^{ss}/e_{Dm} =  0.30 $, where $e_{Dm}$ is $e_D$ evaluated approximately at metling temperature of copper i.e., at $T_k \approx 1350$ K. $\hat{c}_ce_{Dm}= \bar{\mu}_T/14$ at 298 K, $\bar{\mu}_T/100$ at 1023 K and $\bar{\mu}_T/180$ at 1173 K. The experimental data reported here are taken from \citet{preston2003model} and \citet{voyiadjis2005microstructural}.

As evident from Fig. \ref{stress_strain_high_temp}, the model prediction matches fairly well with experimental observations in the high temperature (1023 K and 1173 K) regime. However, for low strain rates (0.066/sec), the match is comparatively inferior to that for high strain rates (960/sec and 1800/sec). Fig. \ref{stress_strain_room_temp} presents model prediction and experimental data at room temperature (298 K) at moderate (451/sec) as well as high (8000/sec) strain rates. It is clear that the model prediction at room temperature is not as good as that for high temperature. This shortcoming at low temperature and low strain rates may be attributed to the one internal variable modelling of the configurational subsystem, i.e. using a sole dislocation density. Indeed, at low temperature and low strain rates, the significant contribution of immobile dislocations in determining the plastic response is not properly reflected in this model. It, accordingly, does establish the need for an extension to include multiple dislocation densities as described in section \ref{sec_extension}.

\begin{figure}[!htp]
\begin{center}
\includegraphics[width=0.7\textwidth]{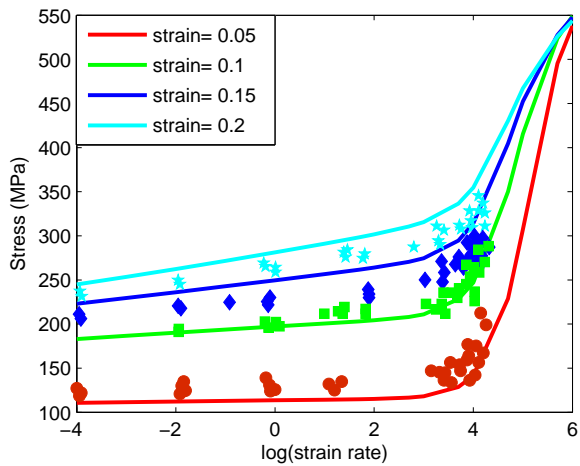}
\end{center}
\caption{Stress vs. plastic strain rate at room temperature}%
\label{strain_rate_low_temp}
\end{figure}

In order to assess the merit of the model in predicting the experimentally observed extreme rate sensitivity of the flow stress near the critical strain rate of $10^4\text{/sec}$, we carried out a series of simulations, at room temperature and with different strain rates. Fig. \ref{strain_rate_low_temp} presents the results in form of plots of stress vs. logarithm of strain rates for given strains. These results conform well with experimental observations and indeed capture the sharp upturn of the flow stress near the critical strain rate. We may expect a still improved quantitative match with the experimental data when the present model is extended to feature multiple dislocation densities.

\section{Concluding remarks}\label{sec_conclusion}
 En route to a physical theory for metal viscoplasticity, a two-temperature thermodynamic setup has been adopted wherein constraints on the material constitution follow from a variant of the fluctuation relation that supersedes and generalizes the second law and helps making the constraints sharper. Starting with an idealised homogeneous setup, a full fledged three dimensional continuum model for elasto-viscoplasticity in a yield-free format is arrived at. The yield-free construction, based on a microscopic force balance that supplements the usual macroscopic force balance, dispenses with the conventional flow rule and loading-unloading conditions. The plastic deformation state is tracked physically through the dislocation density evolution.

 This is work in progress and has its limitations. The model, for instance, considers a sole dislocation density, not distinguishing between its mobile/sessile or statistically stored/geometrically necessary components. Being focussed only on FCC metals in developing the constitutive relations, an extension to other metal types is also a matter of future work. The most significant extension, though, would likely be around a reinterpretation of the fluctuation relation, entailing a characterization of the entropy exponent as an exponential martingale whose first moment (Eq. \eqref{eq_fluc3}) is only used in this work.

\section*{Reference}
\bibliographystyle{elsarticle-harv}
\bibliography{ref}

\end{document}